\documentclass[12pt]{iopart}

\usepackage{iopams}
\usepackage{graphicx}
\usepackage{mathrsfs}
\usepackage{color}
\usepackage[table]{xcolor}
\usepackage{hyperref}
\hypersetup{colorlinks, citecolor=black, filecolor=black, linkcolor=black, urlcolor=black}

\expandafter\let\csname equation*\endcsname\relax
\expandafter\let\csname endequation*\endcsname\relax
\usepackage{amsmath}

\definecolor{grayish}{RGB}{230,230,230}

\newcommand{\refEq}[1] {(\ref{#1})}

\begin{document}

\title{Intuition for the radial penetration of flux surface shaping in tokamaks}

\author{Justin Ball and Felix I. Parra}

\address{Rudolf Peierls Centre for Theoretical Physics, University of Oxford, Oxford OX1 3NP, United Kingdom}
\address{Culham Centre for Fusion Energy, Culham Science Centre, Abingdon OX14 3DB, United Kingdom}
\ead{Justin.Ball@physics.ox.ac.uk}

\begin{abstract}

Using analytic calculations, the effects of the edge flux surface shape and the toroidal current profile on the penetration of flux surface shaping are investigated in a tokamak. It is shown that the penetration of shaping is determined by the poloidal variation of the poloidal magnetic field on the surface. This fact is used to investigate how different flux surface shapes penetrate from the edge. Then, a technique to separate the effects of magnetic pressure and tension in the Grad-Shafranov equation is presented and used to calculate radial profiles of strong elongation for nearly constant current profiles. Lastly, it is shown that more hollow toroidal current profiles are significantly better at conveying shaping from the edge to the core.

\end{abstract}



\section{Introduction}
\label{sec:introduction}

Recently, it has been shown theoretically \cite{PeetersMomTransSym2005,ParraUpDownSym2011,SugamaUpDownSym2011}, experimentally \cite{CamenenPRLExp2010}, and numerically \cite{BallMomUpDownAsym2014} that breaking the up-down symmetry of tokamak flux surfaces can significantly increase the intrinsic toroidal rotation without the need for an external momentum source. Toroidal rotation has been experimentally proven to increase MHD stability \cite{GarofaloExpRWMstabilizationD3D2002,ReimerdesRWMmachineComp2006,SabbaghExpRWMstabilizationNSTX2002,StraitExpRWMstabilizationD3D1995,deVriesRotMHDStabilization1996} and can suppress turbulent energy transport \cite{BarnesFlowShear2011,BurrellShearTurbStabilization1997,HighcockRotationBifurcation2010,ParraMomentumTransitions2011,RitzRotShearTurbSuppression1990}. This has motivated substantial interest in creating strong up-down asymmetry that penetrates radially into the plasma \cite{BallMomUpDownAsym2014,RodriguesMHDupDownAsym2014,BizarroUpDownAsymGradShafEq2014}. Additionally, strong shaping increases the stability of the plasma to kink modes by increasing the safety factor at constant plasma current \cite{SugiharaITERdisruptionVDE2007}. Also, shaping has been observed to have a stabilizing effect on ELMs \cite{OzekiTurbStableElongation1990} and to improve turbulent energy transport \cite{WaltzITGstabilElongation1999,ShimadaITER2007}.

This work uses a series of independent arguments to develop a more general intuition for existing analytic \cite{LaoGradShafExpansion1981,CerfonSolovevGradShafSols2010,AtanasiuGradShafSols2004} and numerical \cite{RomanelliShapingPenetration2000,LaoShapeAndCurrent1985} results concerning how plasma shaping penetrates in the ideal MHD model \cite{FreidbergIdealMHD1987}. We investigate the effects of both free parameters in the Grad-Shafranov equation \cite{GradGradShafranovEq1958}: the boundary condition and the toroidal current profile. Although this work was motivated by up-down asymmetry, the main results of this paper apply to the penetration of traditional up-down symmetric plasma shaping as well. The following derivations are appropriate to treat the Shafranov shift, however it will not be investigated specifically. This is because, in isolation, it does not create up-down asymmetry and is formally small in aspect ratio. However, it is worth noting that the Shafranov shift becomes up-down asymmetric when the flux surfaces already have an up-down asymmetric shape. Hence it can enhance existing up-down asymmetry, but cannot create asymmetry by itself.

The traditional argument concerning shaping penetration \cite{RodriguesMHDupDownAsym2014,BizarroUpDownAsymGradShafEq2014,LaoGradShafExpansion1981} uses a Taylor expansion of the poloidal flux function about the magnetic axis to find
\begin{eqnarray}
   \psi \left( R, Z \right) \approx& \left. \frac{\partial^{2} \psi}{\partial R^{2}} \right|_{R_{0},0} \left( R - R_{0} \right)^{2} + \left. \frac{\partial^{2} \psi}{\partial R \partial Z} \right|_{R_{0},0} \left( R - R_{0} \right) Z \label{eq:TaylorExpansion} \\
   &+  \left. \frac{\partial^{2} \psi}{\partial Z^{2}} \right|_{R_{0},0} Z^{2} , \nonumber
\end{eqnarray}
where $R$ is the major radial coordinate, $R_{0}$ is the major radial location of the magnetic axis, $Z$ is the vertical coordinate, and we have assumed that the magnetic axis is located at $Z = 0$. We have imposed that the flux vanishes on the magnetic axis and since the flux is at a minimum at the magnetic axis, the linear term is zero. This means that, no matter what external fields shape the plasma, close enough to the magnetic axis, the flux surface ellipticity will dominate over higher order shaping effects. Note that if all the second order Taylor coefficients are zero, then this argument fails. However, this requires a vanishing toroidal current density on-axis, which prevents closed, nested flux surfaces \cite{RodriguesGradShafVanishingCurrent2013}. While this argument is compelling, it says nothing about how shaping behaves away from the magnetic axis or how triangularity penetrates in the absence of elongation. A more sophisticated version of this argument is presented in references \cite{BallMomUpDownAsym2014,RodriguesMHDupDownAsym2014}, which includes effects from having a linear toroidal current profile about the magnetic axis.

In section \ref{sec:calculation}, we show that the shaping of a given flux surface depends on the strength of the poloidal variation of the poloidal magnetic field on the flux surface. Then, in section \ref{sec:fluxSurfaceShapeEffect}, we use this dependence to study why different flux surface shapes penetrate better than others. In section \ref{sec:effectOfCurrentProfile}, we explore a limit of the Grad-Shafranov equation to separate the effects of magnetic pressure and tension. In this limit we clearly see how the current profile affects shaping penetration.

\section{Quantifying shaping penetration}
\label{sec:calculation}

The amount of flux surface shaping can be quantified by the parameter
\begin{eqnarray}
   \Delta \left( a \right) \equiv \frac{b \left( a \right)}{a} ,
\end{eqnarray}
where the flux surface label $a$ is the minor radius (i.e. the minimum distance of the flux surface from the magnetic axis) and $b$ is the maximum distance of the flux surface from the magnetic axis. For circular flux surfaces without a Shafranov shift $\Delta = 1$. Since the definitions of $a$ and $b$ are based on the magnetic axis, $\Delta \neq 1$ for circular flux surfaces with a Shafranov shift. For purely elliptical flux surfaces without a Shafranov shift, $\Delta$ reduces to the typical definition of the elongation, usually denoted by $\kappa$. Experimentally $\Delta$ is typically dominated by the effect of elongation, which can range from $\sim 1$ in unshaped devices to the record value of $2.8$ in TCV \cite{HofmannTCVRecordElong2002}. Taking a derivative we find the change in elongation across a flux surface is given by
\begin{eqnarray}
   \frac{d \Delta}{da} = \frac{1}{a} \frac{db}{da} - \frac{b \left( a \right)}{a^{2}} . \label{eq:kappaDeriv}
\end{eqnarray}
The derivative can be calculated from the poloidal magnetic flux,
\begin{eqnarray}
   \psi = \frac{1}{2 \pi} \oint_{-\pi}^{\pi} d \zeta \int_{0}^{r} d r' R B_{p} , \label{eq:poloidalFluxDef}
\end{eqnarray}
where $\zeta$ is the toroidal angle, $r$ is the distance from the magnetic axis, and $\vec{B}_{p}$ is the poloidal magnetic field. Equation \refEq{eq:poloidalFluxDef} is only valid along the integration path connecting the radial minimum on each flux surface, $a$, and the path connecting the radial maximum on each flux surface, $b$. This is because, at the flux surface radial extrema, the poloidal field is necessarily perpendicular to the usual cylindrical radial direction. Using implicit differentiation and evaluating on both of these integration paths, equation \refEq{eq:poloidalFluxDef} gives
\begin{eqnarray}
   \frac{d a}{d \psi} = \frac{1}{\left. R B_{p} \right|_{a}} , \label{eq:minorRadiusFluxDeriv} \\
   \frac{d b}{d \psi} = \frac{1}{\left. R B_{p} \right|_{b}} . \label{eq:majorRadiusFluxDeriv}
\end{eqnarray}
Here $|_{a}$ and $|_{b}$ indicate the quantity should be evaluated at the poloidal locations of the minimum and maximum radial positions on a given flux surface. Therefore, we find that equation \refEq{eq:kappaDeriv} becomes
\begin{eqnarray}
   \frac{a}{\Delta} \frac{d \Delta}{da} = \frac{1}{\Delta} \frac{ \left. R B_{p} \right|_{a}}{\left. R B_{p} \right|_{b}}  - 1 . \label{eq:kappaDerivFields}
\end{eqnarray}
In current experiments \cite{BrixFluxSurfShapes2008,HofmannTCVRecordElong2002,LaoShapeAndCurrent1985,LaoDIIIDfluxSurfShapes2005} this quantity is generally between $0$ and $0.3$, but, as additional shaping is generally advantageous, the goal would be to make it as negative as possible. We will use equation \refEq{eq:kappaDerivFields} to understand why different flux surface shapes (elongated, triangular, etc.) penetrate better from the edge to the core and how the toroidal current profile affects this penetration.

\section{Effect of flux surface shape}
\label{sec:fluxSurfaceShapeEffect}

\begin{figure}
 \hspace{0.04\textwidth} (a) \hspace{0.255\textwidth} (b) \hspace{0.255\textwidth} (c) \hspace{0.1\textwidth}

 \centering
 \includegraphics[width=0.3\textwidth]{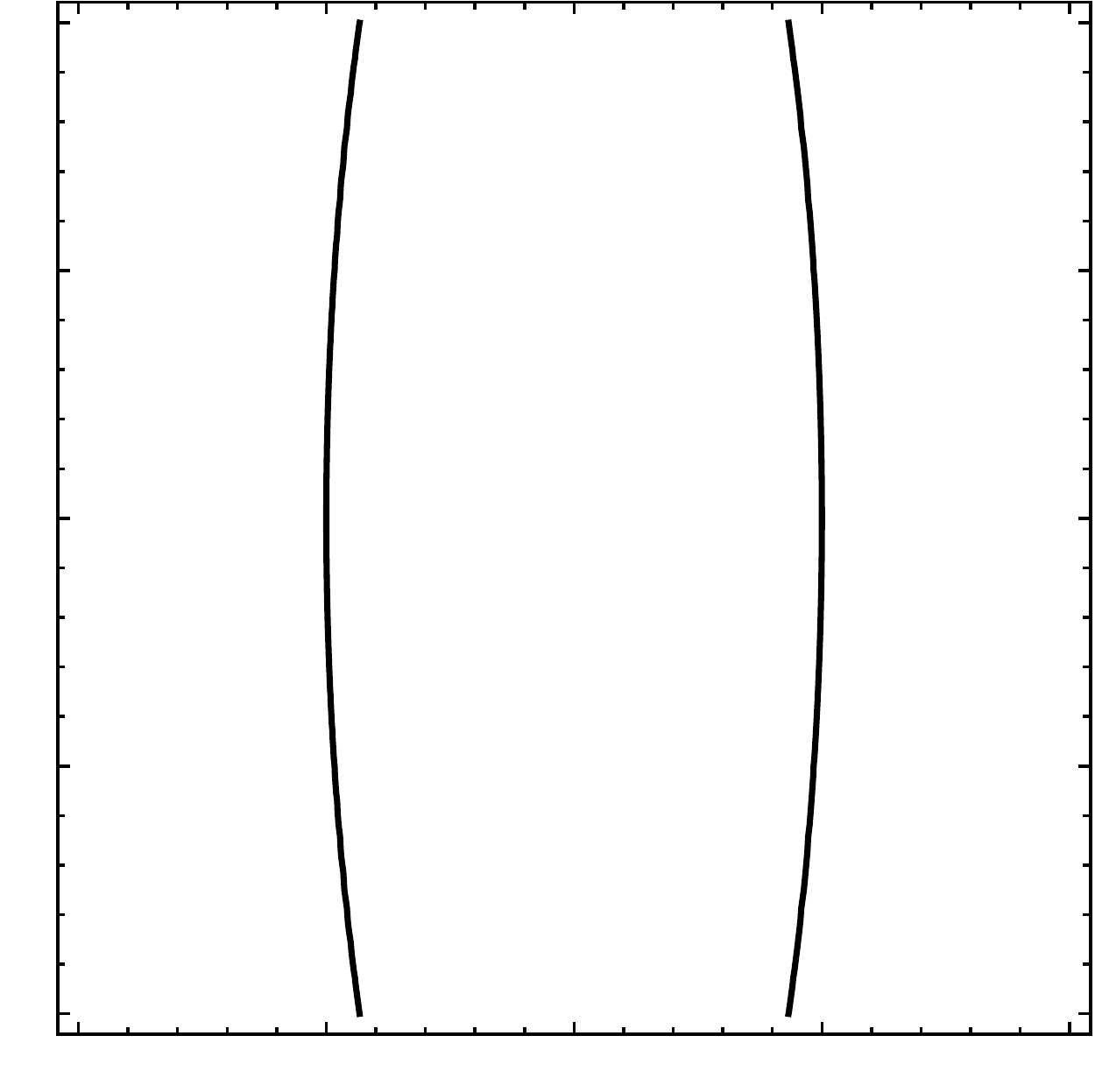}
 \includegraphics[width=0.3\textwidth]{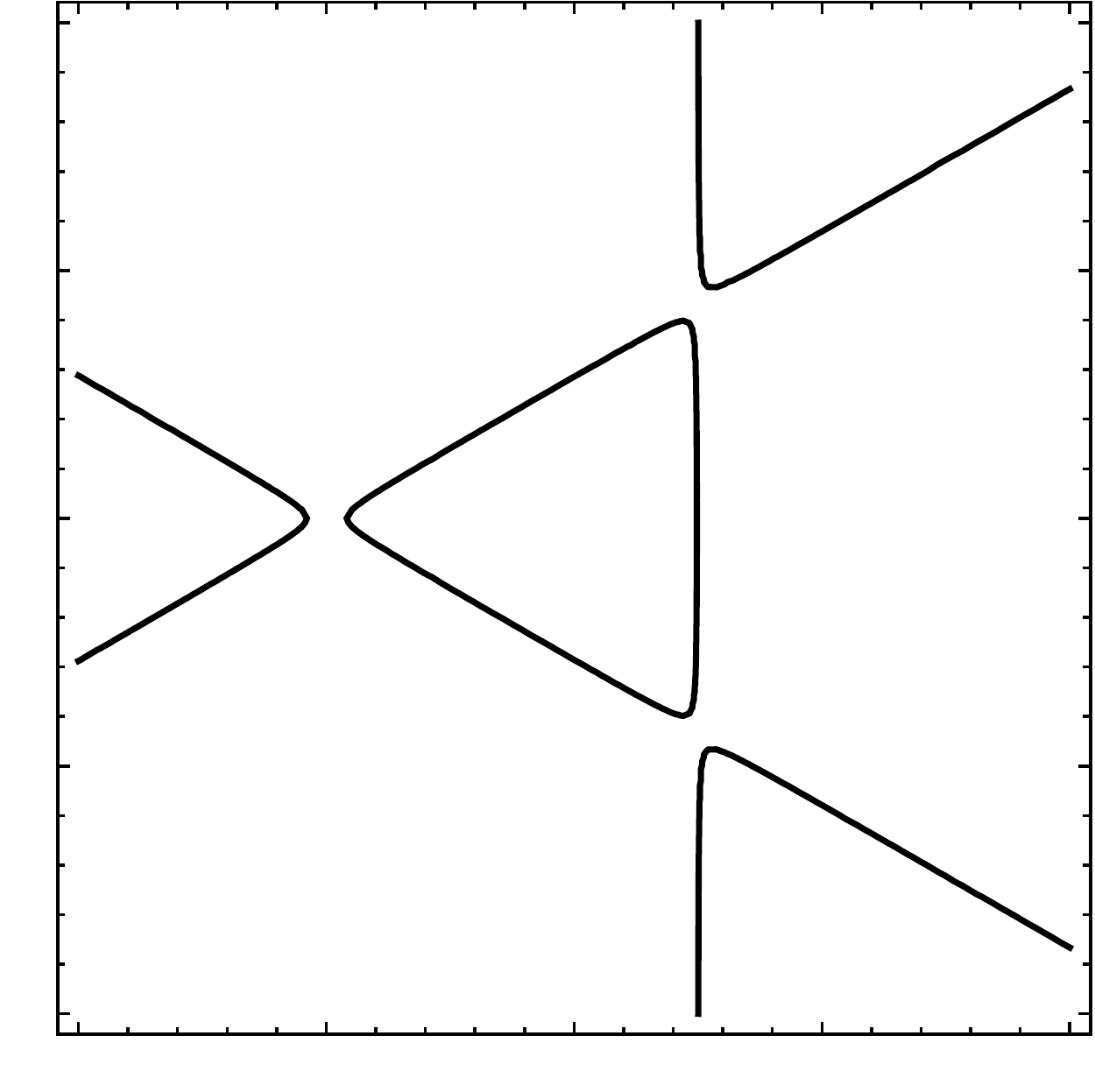}
 \includegraphics[width=0.3\textwidth]{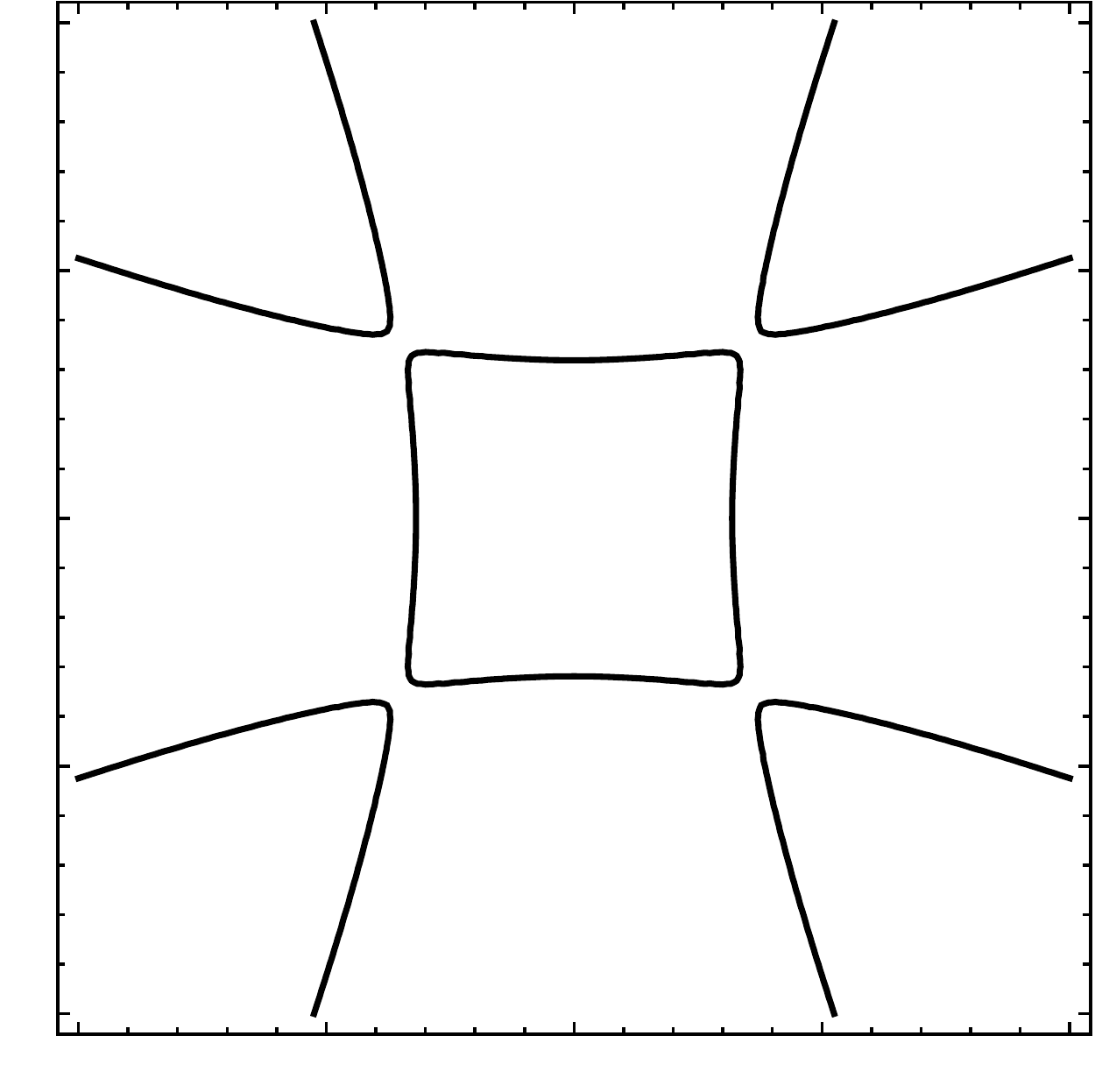}
 \caption{The (a) $m=2$, (b) $m=3$, and (c) $m=4$ strongly shaped flux surface shapes.}
 \label{fig:equilibriaFluxShapes}
\end{figure}

In this section we will compare different flux surface shapes and show that lower order shaping effects penetrate from the plasma boundary to the magnetic axis more effectively. First, we must determine which shapes to consider and argue that comparisons between them are fair. We will use large aspect ratio equilibria produced with a constant toroidal current profile because it is a reasonable approximation of experimental profiles and the solutions are simple cylindrical harmonics. From these equilibria we will investigate each cylindrical harmonic shaping effect in isolation by creating strongly shaped flux surfaces, specifically those that approach having magnetic field nulls (see figure \ref{fig:equilibriaFluxShapes}). These configurations will be analytically tractable and exaggerate the effects we mean to investigate. It should be noted that we expect flux surfaces with higher order shaping to be more difficult to create experimentally. This is because they have more magnetic field nulls, so they require more magnets and more total external current to create.

From Ampere's law we find that $\left. R B_{p} \right|_{a} \approx \left(S_{p} / l_{p} \right) \mu_{0} j_{\zeta} R$, where $S_{p}$ is the poloidal area enclosed by the flux surface, $l_{p}$ is the poloidal perimeter, $\mu_{0}$ is the vacuum permeability, and $j_{\zeta}$ is the toroidal current. Crucially, we note that $\left. R B_{p} \right|_{b}$ is small because we have chosen configurations that nearly have magnetic nulls. What this reveals is, as the flux surface shaping is increased, the ratio of poloidal fields in equation \refEq{eq:kappaDerivFields} diverges to positive infinity. This implies that $d \Delta / d a$ is positive and large, that is, for a reasonably physical current profile, it will be impossible to maintain strong shaping from the boundary to the magnetic axis. While this is true for nearly all configurations, there is one caveat: when the shaping parameter $\Delta$ also diverges to infinity. Then, $d \Delta / d a$ can be finite and negative. This makes the $m=2$ cylindrical harmonic shaping effect special because flux surfaces with arbitrarily large elongation are possible. All pure higher order shaping effects cannot make flux surfaces that are both closed and have arbitrarily large shaping.

When the mathematics are worked out rigorously \cite{BallMomUpDownAsym2014,RodriguesMHDupDownAsym2014,ChristiansenCurrentDist1982} we find that, to lowest order in aspect ratio, a constant current profile has no effect on the externally applied elongation, meaning that $d \Delta / da = 0$. This section has shown that elongation is optimal for radial penetration. In the next section we will investigate the effect of the toroidal current profile on flux surface elongation.

\section{Effect of toroidal current profile}
\label{sec:effectOfCurrentProfile}

As we compare configurations with different toroidal current profiles, we will choose to keep the external flux surface shape fixed. Therefore, from equation \refEq{eq:kappaDerivFields} we conclude that changing the current profile, while maintaining a constant boundary flux surface shape, only affects the shaping penetration by altering $\left. R B_{p} \right|_{a} / \left. R B_{p} \right|_{b}$.

In order to calculate the ratio of the poloidal fields we will start with the toroidal component of Ampere's law,
\begin{eqnarray}
   \left( \vec{\nabla} \times \vec{B} \right) \cdot \hat{e}_{\zeta} = \mu_{0} j_{\zeta} .
\end{eqnarray}
Noting that $\vec{B} = I \vec{\nabla} \zeta + \vec{B}_{p}$, we see that
\begin{eqnarray}
  \left( \vec{\nabla} \times \vec{B}_{p} \right) \cdot R \vec{\nabla} \zeta = \mu_{0} j_{\zeta} .
\end{eqnarray}
Since $\vec{B}_{p} = \vec{\nabla} \zeta \times \vec{\nabla} \psi$, we know that $\vec{\nabla} \zeta = \vec{\nabla} \psi \times \vec{B}_{p} / \left| \vec{\nabla} \psi \right|^{2}$. Making this substitution and using a number of vector identities on the quantity $\vec{B}_{p} \times \left( \vec{\nabla} \times \vec{B}_{p} \right)$ we find that
\begin{eqnarray}
   R \frac{\vec{\nabla} \psi}{\left| \vec{\nabla} \psi \right|^{2}} \cdot \left( \vec{\nabla} \vec{B}_{p} \right) \cdot \vec{B}_{p} - \frac{R B_{p}^{2}}{\left| \vec{\nabla} \psi \right|^{2}} \hat{b}_{p} \cdot \left( \vec{\nabla} \hat{b}_{p} \right) \cdot \vec{\nabla} \psi = \mu_{0} j_{\zeta} ,
\end{eqnarray}
where $\hat{b}_{p} \equiv \vec{B}_{p} / B_{p}$ is the poloidal field unit vector. Using the definition of the poloidal field curvature vector, $\vec{\kappa}_{p} \equiv \hat{b}_{p} \cdot \vec{\nabla} \hat{b}_{p}$, together with $\vec{\nabla} \psi = R B_{p} \hat{e}_{\psi}$ (which are necessarily antiparallel) gives
\begin{eqnarray}
   \frac{R}{2} \frac{\vec{\nabla} \psi}{\left| \vec{\nabla} \psi \right|^{2}} \cdot \vec{\nabla} \left( B_{p}^{2} \right) + B_{p} \kappa_{p} = \mu_{0} j_{\zeta} . \label{eq:gradShafranovSimple}
\end{eqnarray}
We choose this form because it clearly separates the effects of poloidal magnetic pressure in the first term and field line tension in the second, while the right hand side is constant on a flux surface to lowest order in aspect ratio. Equation \refEq{eq:gradShafranovSimple} is a different way to express the conclusion reached in reference \cite{ChristiansenCurrentDist1982}: in non-circular flux surfaces, the current profile determines the shaping. In equation \refEq{eq:gradShafranovSimple}, this result is given in terms of the poloidal magnetic field, which can then be related to the shaping with equation \refEq{eq:kappaDerivFields}.

\begin{figure}
 \hspace{0.04\textwidth} (a) \hspace{0.255\textwidth} (b) \hspace{0.255\textwidth} (c) \hspace{0.1\textwidth}
 \begin{center}
  \includegraphics[width=0.3\textwidth]{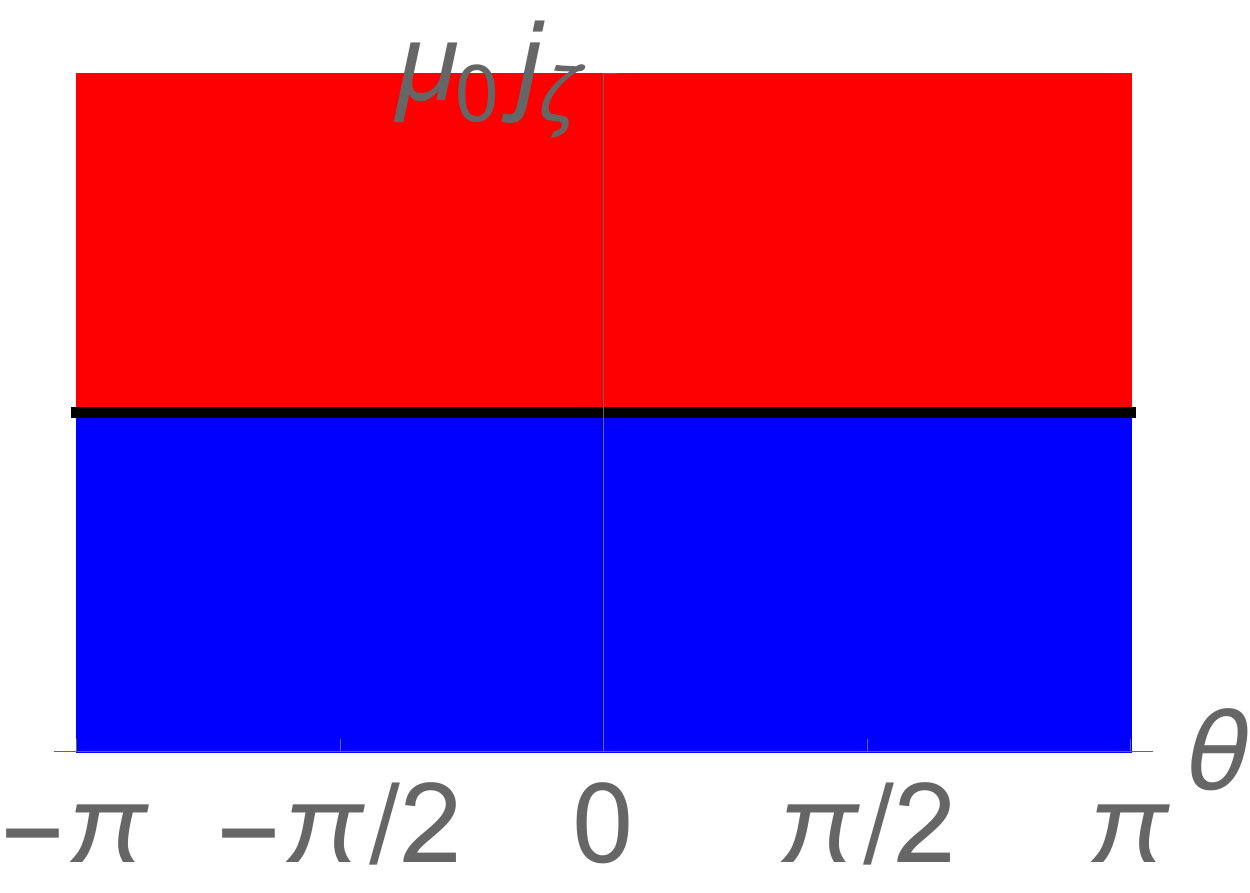}
  \includegraphics[width=0.3\textwidth]{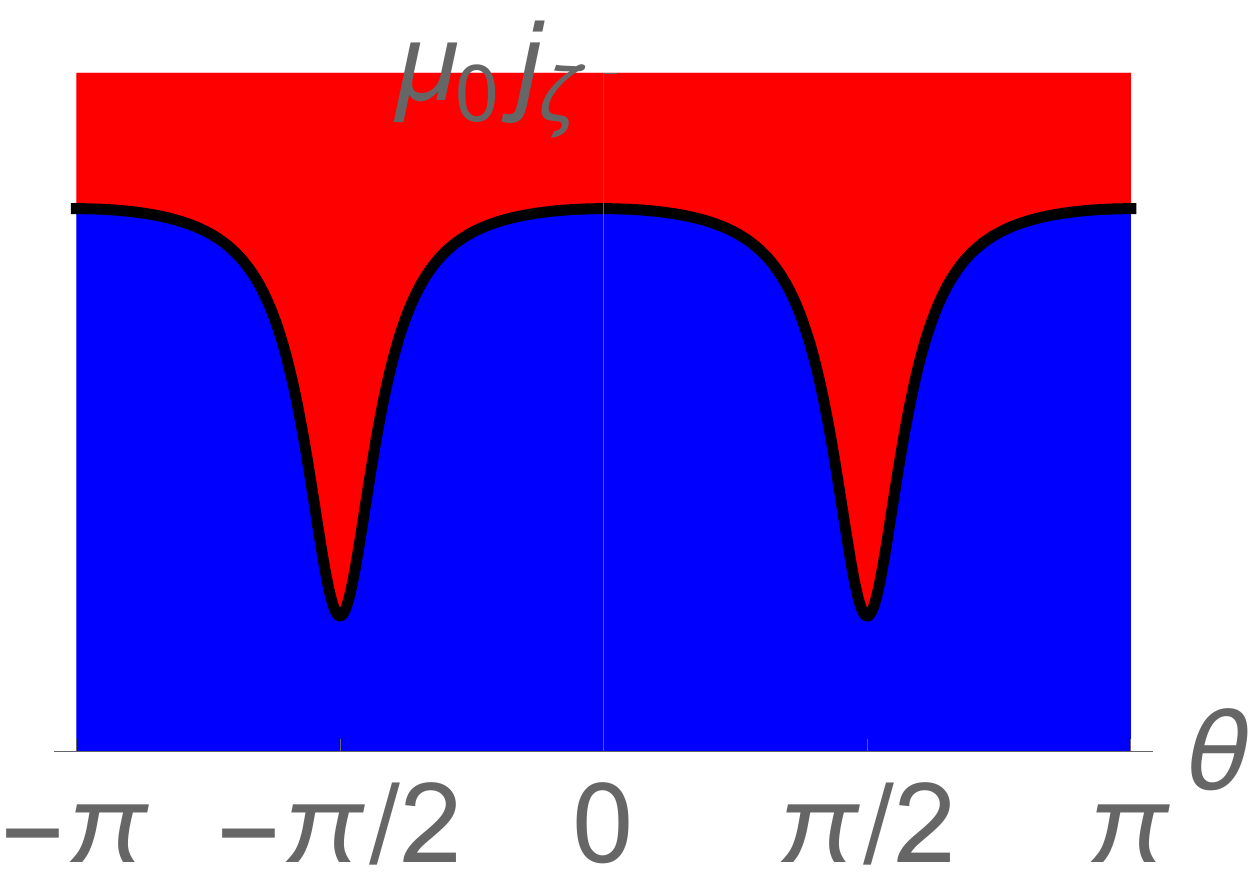}
  \includegraphics[width=0.3\textwidth]{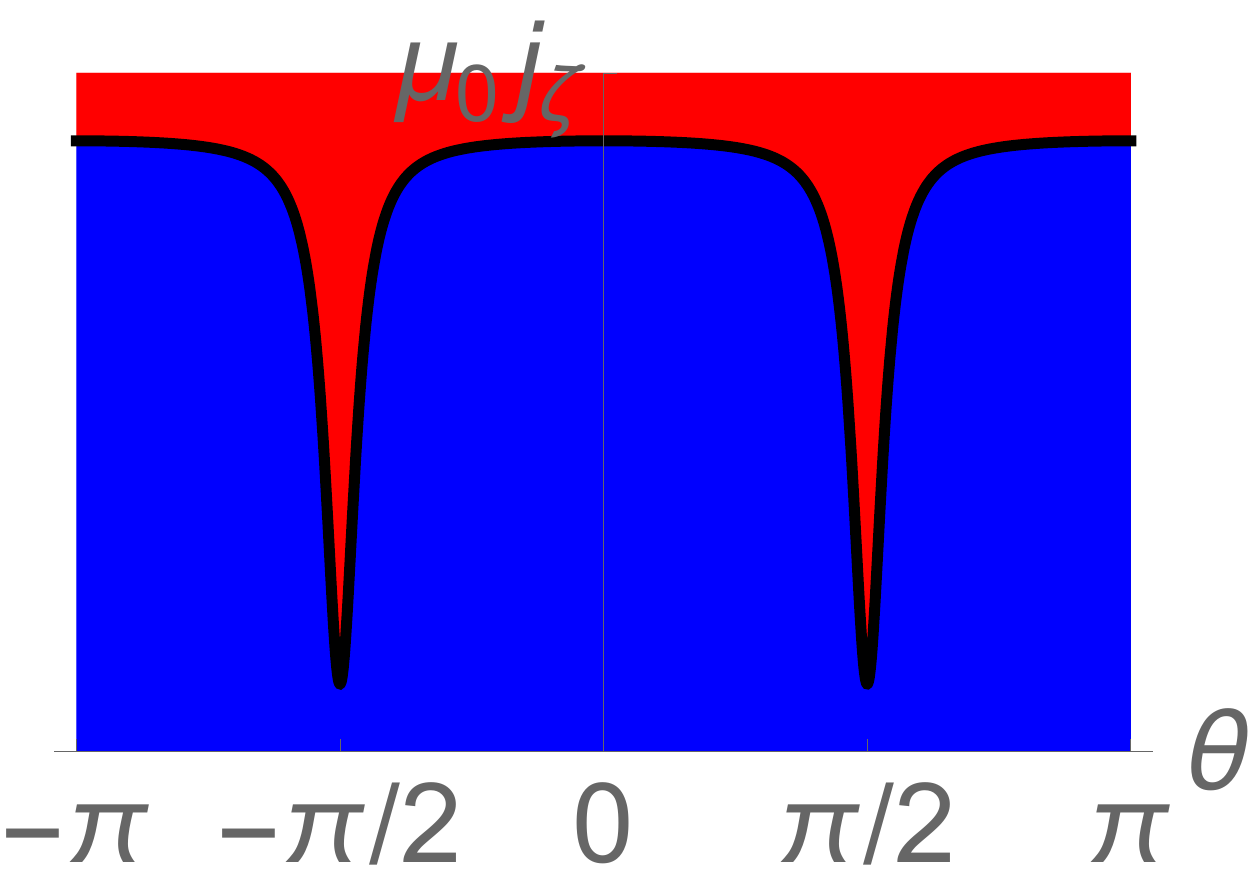}
 \end{center}
 \caption{A stacked area graph showing, to lowest order in aspect ratio, the contributions of the magnetic pressure (blue, below the curve) and tension (red, above the curve) terms from equation \refEq{eq:gradShafranovSimple} on an elongated flux surface with (a) $\Delta = 1$, (b) $\Delta = 2$, and (c) $\Delta = 3$, where $\theta$ is the traditional cylindrical poloidal angle.}
 \label{fig:pressureTensionBalance}
\end{figure}

We apply equation \refEq{eq:gradShafranovSimple} to strongly shaped flux surfaces, which causes the first and second terms to vary dramatically with the poloidal location. We will assume that, at the poloidal location of the minimum radial position, the field lines become straight and the curvature term vanishes. Additionally, since the poloidal derivative necessarily vanishes at this location, the gradient can be converted according to the chain rule as
\begin{eqnarray}
   \left. \vec{\nabla} \left( B_{p}^{2} \right) \right|_{a} = \left. \vec{\nabla} \psi \right|_{a} \frac{d a}{d \psi} \frac{d}{d a}  \left( \left. B_{p}^{2} \right|_{a} \right) .
\end{eqnarray}
Then equation \refEq{eq:minorRadiusFluxDeriv} can be used to find
\begin{eqnarray}
   \left. B_{p} \right|_{a} = \mu_{0} \int_{0}^{a} d a' \left. j_{\zeta} \right|_{a} \left( a' \right) . \label{eq:gradShafranovPressure}
\end{eqnarray}
Furthermore, we assume that, at the poloidal location of the maximum radial position, the magnetic pressure term is small, giving
\begin{eqnarray}
   \left. B_{p} \right|_{b} = \left. \frac{\mu_{0} j_{\zeta}}{\kappa_{p}} \right|_{b} . \label{eq:gradShafranovTension}
\end{eqnarray}
The integral in equation \refEq{eq:gradShafranovPressure} assumes that the separation between magnetic pressure and tension must be valid over the entire radial profile, not just on the flux surface of interest. If the flux surfaces are circular over a substantial region near the axis, equation \refEq{eq:gradShafranovPressure} is no longer accurate. For the $m=2$ mode with a constant current profile, equations \refEq{eq:gradShafranovPressure} and \refEq{eq:gradShafranovTension} are exact in the limits of $\Delta \rightarrow \infty$ and $a/R \rightarrow 0$ (see figure \ref{fig:pressureTensionBalance}). This is because, in these conditions, the flux surface exactly maintains its shape as it penetrates the plasma \cite{BallMomUpDownAsym2014,RodriguesMHDupDownAsym2014}. With a linear peaked current profile that changes by 20\% over the radial region and an elongation of $\Delta = 2$, equations  \refEq{eq:gradShafranovPressure} and \refEq{eq:gradShafranovTension} are only accurate to about 20\%. These equations are not exact for other types of shaping, but we will keep the derivation completely general because approximate results may still be useful and other exact limits may exist for different current profiles.

Substituting equations \refEq{eq:gradShafranovPressure} and \refEq{eq:gradShafranovTension} into equation \refEq{eq:kappaDerivFields} we find that
\begin{eqnarray}
   \frac{a}{\Delta} \frac{d \Delta}{da} = \frac{\left. \kappa_{p} \right|_{b}}{\Delta} \frac{\left. R \right|_{a} \int_{0}^{a} d a' \left. j_{\zeta}\right|_{a} \left( a' \right)}{\left. R \right|_{b} \left. j_{\zeta} \right|_{b}}  - 1 . \label{eq:kappaDerivCurrents}
\end{eqnarray}
Since we are considering the flux surface shape as fixed, we can solve for the required current profile properties to locally permit the shape to penetrate (i.e. $d \Delta / d a = 0$) and find
\begin{eqnarray}
   \frac{\left. \kappa_{p} \right|_{b}}{\Delta} = \frac{ \left. R \right|_{b} \left. j_{\zeta c} \right|_{b}}{\left. R \right|_{a} \int_{0}^{a} d a' \left. j_{\zeta c} \right|_{a} \left( a' \right)} . \label{eq:geometricQuantites}
\end{eqnarray}
Here $j_{\zeta c}$ is the toroidal current density profile required for $d \Delta / d a = 0$ locally. We are guaranteed that a solution to equation \refEq{eq:geometricQuantites} exists for every boundary flux surface shape because, by different choices of $j_{\zeta c}$ we can make the right-hand side span the full range of $\left[ 0, \infty \right)$. Solving for this constant shape penetration case is useful because we are comparing configurations holding the flux surface shape constant, so both $\left. \kappa_{p} \right|_{b}$ and $\Delta$ will stay fixed. Substituting equation \refEq{eq:geometricQuantites} into equation \refEq{eq:kappaDerivCurrents}, we find that
\begin{eqnarray}
   \frac{a}{\Delta} \frac{d \Delta}{da} = \frac{\left. j_{\zeta c} \right|_{b}}{\left. j_{\zeta} \right|_{b}} \frac{ \int_{0}^{a} d a' \left. j_{\zeta} \right|_{a} \left( a' \right)}{\int_{0}^{a} d a' \left. j_{\zeta c} \right|_{a} \left( a' \right)}  - 1 . \label{eq:verificationEq}
\end{eqnarray}
By normalizing this equation, we see that the total plasma current can be scaled without changing the flux surface shapes (by scaling the external currents accordingly). In other words, we can multiply $j_{\zeta c}$ or $j_{\zeta}$ by any numerical factor without changing any flux surface shapes. Equation \refEq{eq:verificationEq} is a differential equation for $\Delta \left( a \right)$, which can be solved giving
\begin{eqnarray}
   \frac{\Delta \left( a \right)}{\Delta_{\text{edge}}} = \text{exp} &\left( -\int_{a}^{a_{\text{edge}}} d a' \left( \frac{1}{a'} \left( \frac{\left. j_{\zeta c} \right|_{b} \left( a' \right)}{\left. j_{\zeta} \right|_{b} \left( a' \right)} \frac{\int_{0}^{a'} d a'' \left. j_{\zeta} \right|_{a} \left( a'' \right)}{\int_{0}^{a'} d a'' \left. j_{\zeta c} \right|_{a} \left( a'' \right)} - 1 \right) \right) \right) , \label{eq:exactShapingProfile}
\end{eqnarray}
where $\Delta_{\text{edge}}$ is the shaping parameter of the outermost flux surface and $a_{\text{edge}}$ is the minor radius of the outermost flux surface.

\begin{figure}
 \centering
 \includegraphics[width=0.8\textwidth]{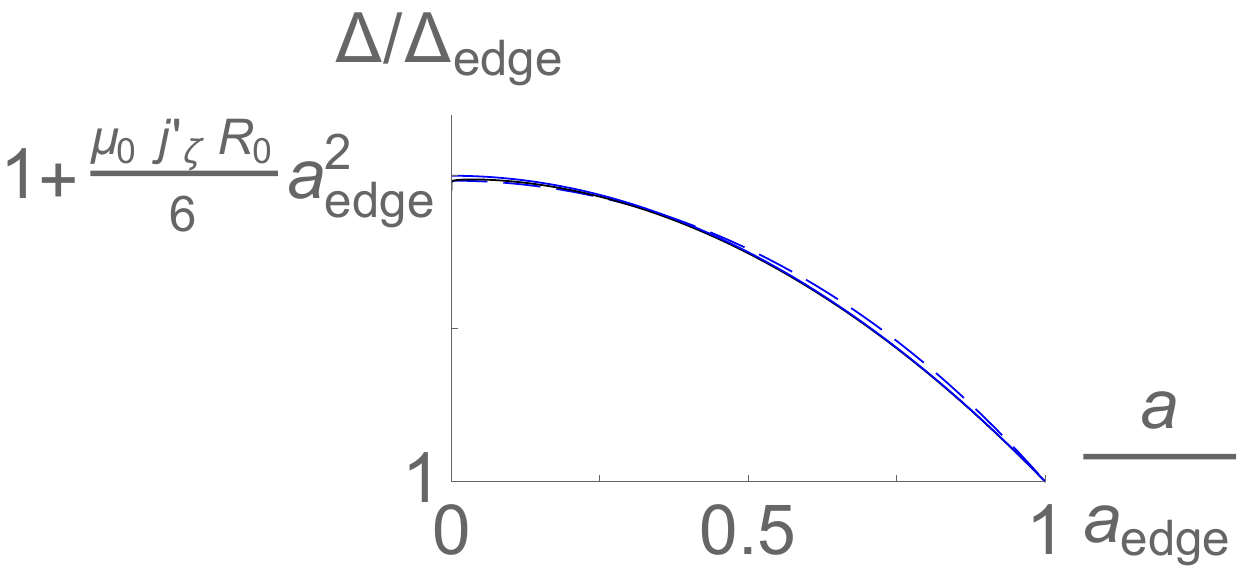}
 \caption{The exact radial shaping profile (black, solid) along with equation \refEq{eq:exactShapingProfile} (blue, dashed) and equation \refEq{eq:quadraticShapingProfile} (blue, solid), which are nearly indistinguishable, for elongated flux surfaces in the limit that $j_{\zeta}' \rightarrow 0$, $a/R \rightarrow 0$, and $\Delta_{\text{edge}} \rightarrow \infty$.}
 \label{fig:limitVerification}
\end{figure}

This equation gives the radial profile of the flux surface shaping, but it is only exact when the separation of the two terms in equation \refEq{eq:gradShafranovSimple} is valid over the entire radial profile. Now we will show an exact solution for elongated flux surfaces with a linear current profile, $j_{\zeta} = j_{\zeta 0} + j_{\zeta}' \psi$, in the limits that  $j_{\zeta}' \rightarrow 0$, $a/R \rightarrow 0$, and $\Delta_{\text{edge}} \rightarrow \infty$. In these limits we can simplify equation \refEq{eq:exactShapingProfile} to
\begin{eqnarray}
   \frac{\Delta \left( a \right)}{\Delta_{\text{edge}}} = 1 + \frac{\mu_{0}  j_{\zeta}' R_{0}}{6} a_{edge}^{2} \left( 1 - \left( \frac{a}{a_{edge}} \right)^{2} \right) . \label{eq:quadraticShapingProfile}
\end{eqnarray}
Figure \ref{fig:limitVerification} shows good agreement between this simple quadratic profile, equation \refEq{eq:exactShapingProfile}, and the exact numerical solution.

\begin{figure}
 \hspace{0.065\textwidth} (a) \hspace{0.255\textwidth} (b) \hspace{0.23\textwidth} (c) \hspace{0.1\textwidth}
 \begin{center}
  \includegraphics[width=0.35\textwidth]{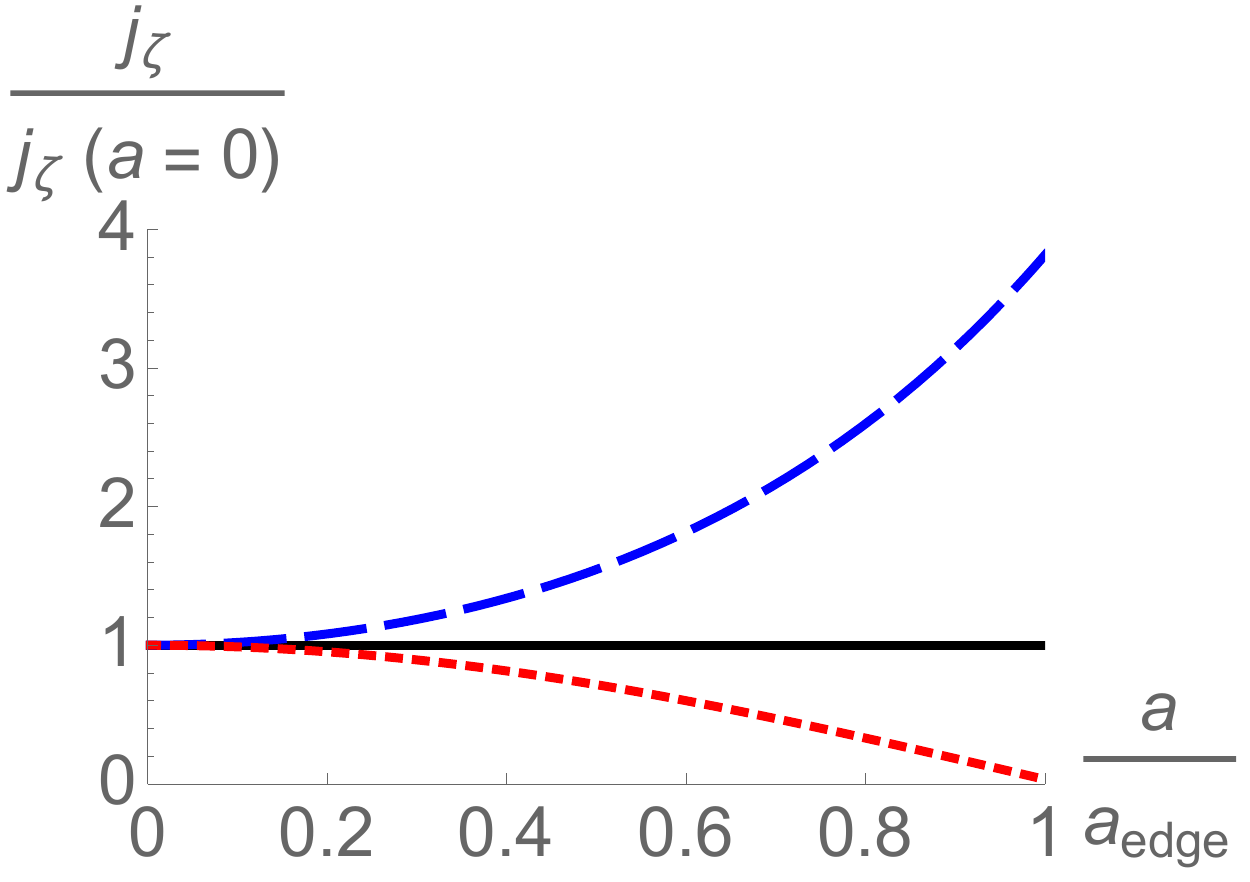}
  \includegraphics[width=0.25\textwidth]{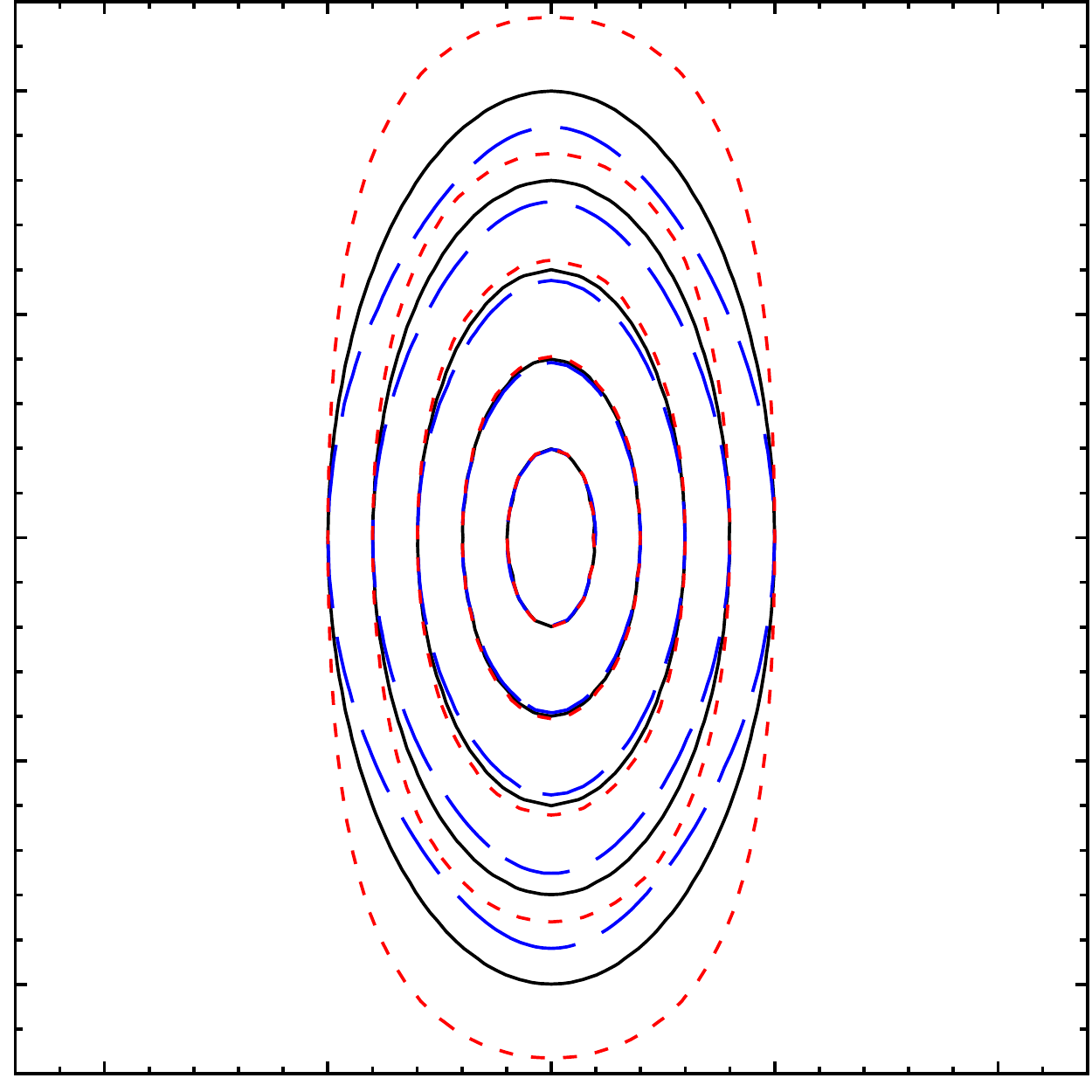}
  \includegraphics[width=0.35\textwidth]{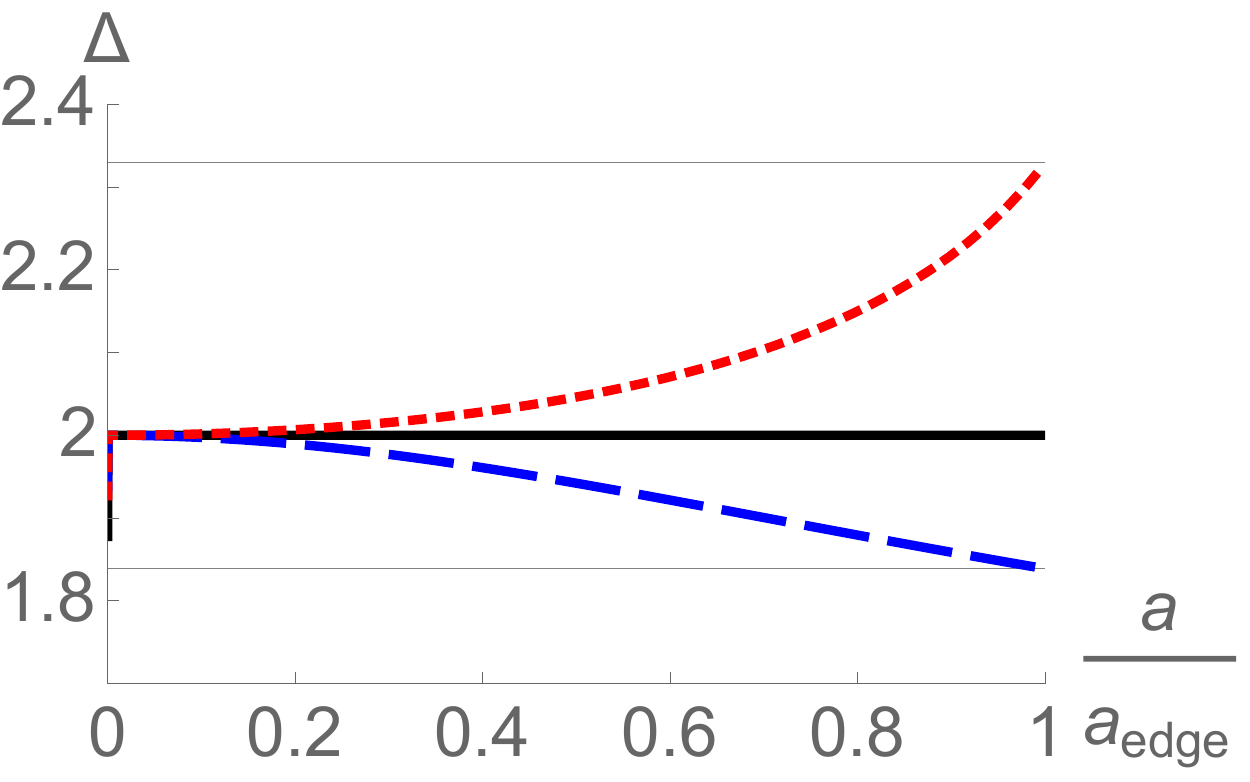}
 \end{center}

 \hspace{0.065\textwidth} (d) \hspace{0.255\textwidth} (e) \hspace{0.23\textwidth} (f) \hspace{0.1\textwidth}
 \begin{center}
  \includegraphics[width=0.35\textwidth]{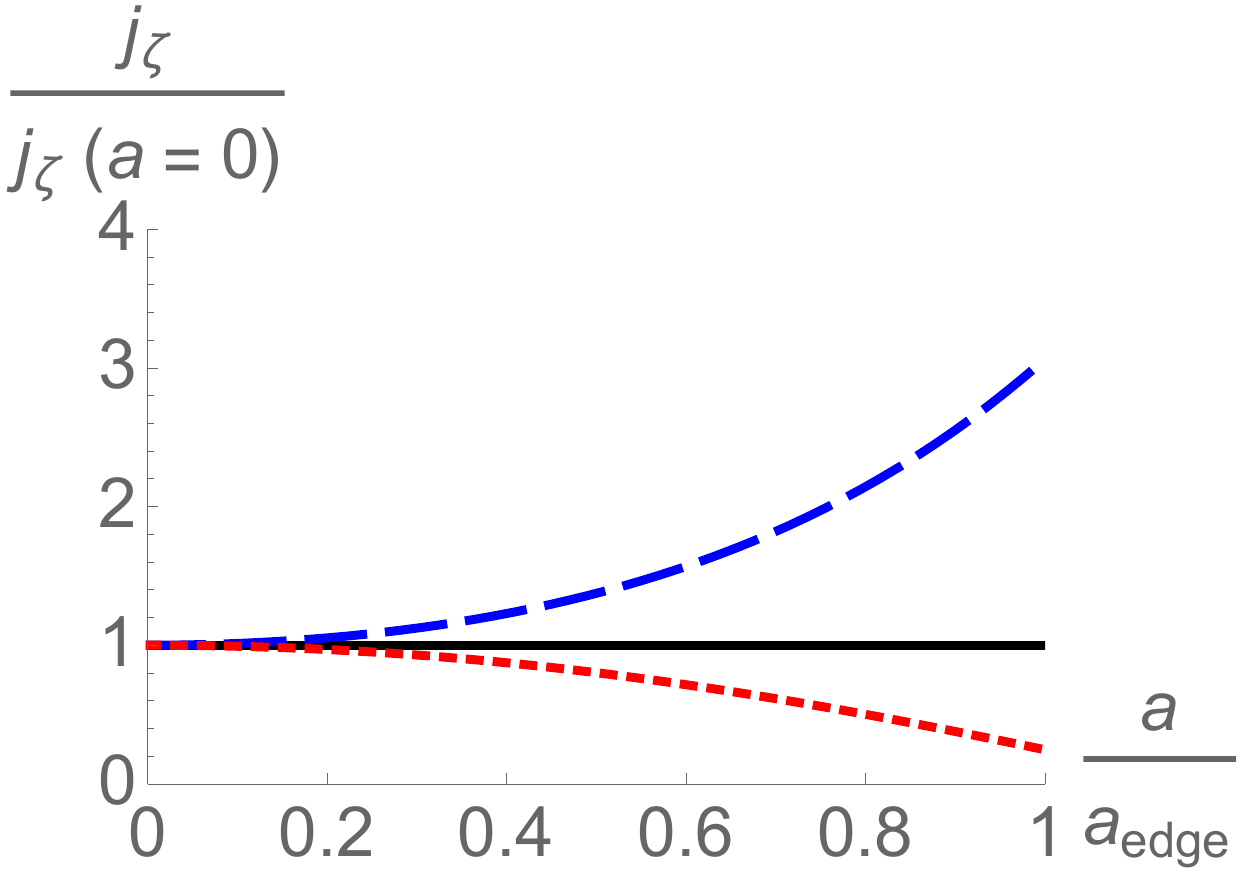}
  \includegraphics[width=0.25\textwidth]{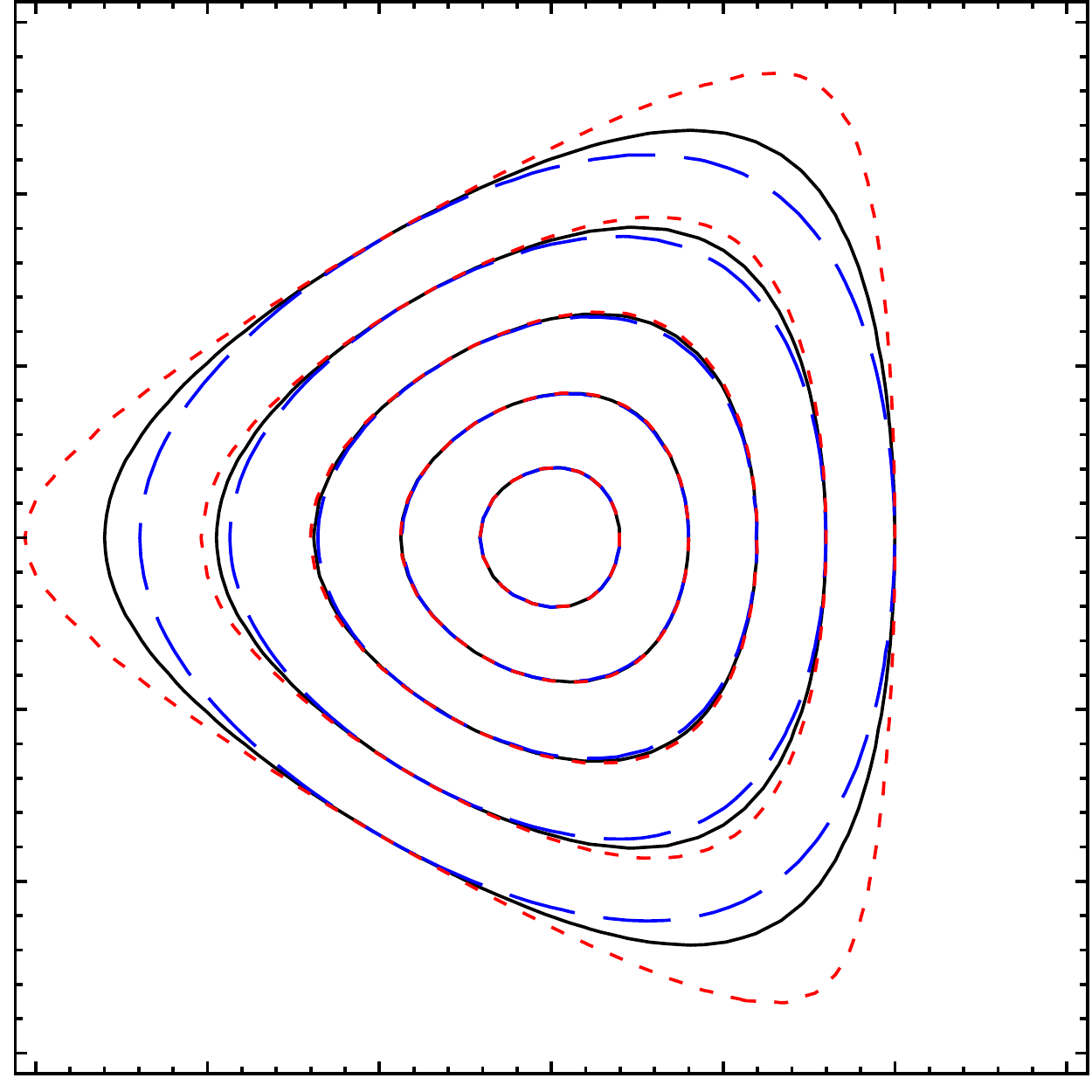}
  \includegraphics[width=0.35\textwidth]{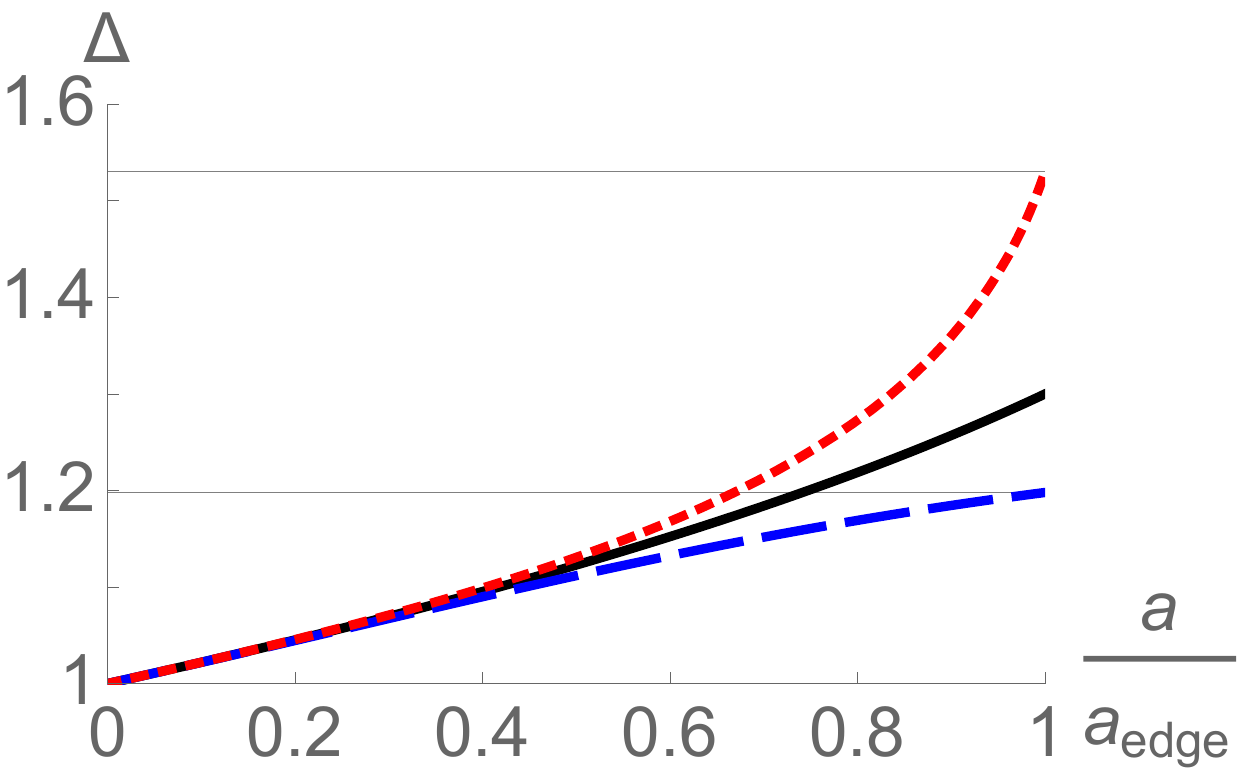}
 \end{center}
 \caption{The (a,d) normalized radial current profile, (b,e) flux surface shapes, and (c,f) shaping profile for solutions to the Grad-Shafranov equation to lowest order in aspect ratio with constant (black, solid), hollow (blue, dashed), and peaked (red, dotted) toroidal current profiles with (a,b,c) elongated or (d,e,f) triangular boundary conditions.}
 \label{fig:exShaping}
\end{figure}

Since $j_{\zeta c}$ can be scaled arbitrarily, equation \refEq{eq:verificationEq} can be further simplified by choosing $\left. j_{\zeta c} \right|_{b}$ to be $\left. j_{\zeta} \right|_{b}$, the toroidal current on the flux surface of interest, giving
\begin{eqnarray}
   \frac{a}{\Delta} \frac{d \Delta}{da} = \frac{\int_{0}^{a} d a' \left. j_{\zeta} \right|_{a} \left( a' \right)}{\int_{0}^{a} d a' \left. j_{\zeta c} \right|_{a} \left( a' \right)}  - 1 \label{eq:kappaDerivFinal}
\end{eqnarray}
at a specific radial location. This shows that the shaping penetration only depends on the amount of toroidal current within the flux surface compared with the constant shape penetration case. Profiles that are more hollow will help shaping penetrate into the plasma. This is because, by definition of ``more hollow,'' $\left. j_{\zeta} \right|_{a}$ is less than $\left. j_{\zeta c} \right|_{a}$. What happens is, as the on-axis current is lowered, the shaping and $\left. R B_{p} \right|_{b}$ stay constant, maintained by the external magnets, while $\left. R B_{p} \right|_{a}$ decreases because of the drop in the total plasma current. From equation \refEq{eq:kappaDerivFields} we see that a change in the ratio of these magnetic fields allows the shaping to penetrate radially. Analogously, peaked current profiles will tend to limit the shaping to the edge. From figure \ref{fig:exShaping}(a,b,c), we see that achieving an on-axis elongation of 2 with a peaked current profile requires a 25\% greater edge elongation than it would with a hollow profile. Figure \ref{fig:exShaping}(d,e,f) shows that triangular flux surface shaping is only large near the boundary, as would be expected from the arguments in sections \ref{sec:introduction} and \ref{sec:fluxSurfaceShapeEffect}. However, we still observe that the shaping penetrates more effectively with a hollow current profile, relative to a peaked profile. This, along with equation \refEq{eq:kappaDerivFinal}, suggests that the beneficial effect of hollow current profiles for shaping penetration is general to all flux surface shapes (see reference \cite{BallMomUpDownAsym2014,RodriguesMHDupDownAsym2014} for a different approach to the same problem). Numerical evidence of this using EFIT equilibrium reconstruction on simulated experimental data can be seen in figure 5(b) of reference \cite{LaoShapeAndCurrent1985}.

\section{Conclusions}
\label{sec:conclusions}

There are several broad points illuminated by this work. First, in section \ref{sec:introduction}, we reviewed the implications of Taylor expanding the poloidal flux about the magnetic axis. It was found that this argument demonstrates that elongation will always dominate higher order shaping near the magnetic axis, but does not forbid higher order shaping from effectively penetrating in the absence of elongation. Next, in section \ref{sec:calculation}, we showed that the change in shaping from flux surface to flux surface depends on the ratio of poloidal magnetic fields at different poloidal locations on the flux surface. Then, in section \ref{sec:fluxSurfaceShapeEffect}, we proved that elongation is the only cylindrical harmonic that can penetrate unaffected from the boundary in the limit of a strongly shaped boundary condition. This suggests that lower order shaping effects always penetrate throughout the plasma most effectively.  Lastly, in section \ref{sec:effectOfCurrentProfile}, we presented a method to separate the effects of magnetic pressure and tension in the Grad-Shafranov equation to get an analytic solution for the shaping penetration of strongly elongated flux surfaces with near constant current profiles. This argument demonstrated hollow current profiles enhance the shaping of strongly elongated elliptical flux surfaces, while peaked current profiles tend to limit elongation to the edge. This effect was able to alter the elongation by over 25\% and appears to be generic to all flux surface shapes.

\ack

The authors would like to thank Paulo Rodrigues and Nuno Loureiro for their helpful suggestions and discussions. This work has been carried out within the framework of the EUROfusion Consortium and has received
funding from the Euratom research and training programme 2014-2018 under grant agreement No. 633053, as well as the RCUK Energy Programme (grant number EP/I501045). The views and opinions expressed herein do not necessarily reflect those of the European Commission.

\section*{References}
\bibliographystyle{unsrt}
\bibliography{references.bib}

\end{document}